\documentclass[showpacs,preprintnumbers,floats,aps,twocolumn]{revtex4}%
\usepackage{graphicx}
\usepackage{amsmath}
\usepackage{amsfonts}
\usepackage{amssymb}%
\providecommand{\U}[1]{\protect\rule{.1in}{.1in}}
\begin{document}
\preprint{ }
\title{Phase sensitive amplification in a superconducting stripline resonator
integrated with a dc-SQUID}
\author{Baleegh Abdo}
\email{baleegh@tx.technion.ac.il}
\author{Oren Suchoi}
\author{Eran Segev}
\author{Oleg Shtempluck}
\author{Eyal Buks}
\affiliation{Department of Electrical Engineering, Technion, Haifa 32000, Israel}
\author{Miles Blencowe}
\affiliation{Department of Physics and Astronomy - Dartmouth College}
\date{\today}

\begin{abstract}
We utilize a superconducting stripline resonator containing a dc-SQUID as a
strong intermodulation amplifier exhibiting a signal gain of 25 dB and a phase
modulation of 30 dB. Studying the system response in the time domain near the
intermodulation amplification threshold reveals a unique noise-induced spikes
behavior. We account for this response qualitatively via solving numerically
the equations of motion for the integrated system. Furthermore, employing this
device as a parametric amplifier yields a gain of 38 dB in the generated
side-band signal.

\end{abstract}

\maketitle




The field of solid-state qubits and quantum information processing has
received considerable attention during the past decade
\cite{Nakamura786,Shnirman357,Makhlin357,Bouchiat165} and has successfully
demonstrated several milestone results to date
\cite{Vion886,Chiorescu1869,Lupascu177006,Wallraff162,Martinis117901,Yamamato823,Claudon187003}%
. However, one of the fundamental challenges hindering this emerging field is
noise interference, which either screens the output signal or leads to quantum
state decoherence
\cite{Vion886,Chiorescu1869,Zorin4408,Sillanpaa066805,Siddiqi207002,Schreier180502}%
. Hence, this may explain to some extent the renewed interest exhibited
recently by the quantum measurement community in the field of phase-sensitive
amplifiers
\cite{Yurke2519,Movshovich1419,Yurke1371,Spietz08062853,Yamamoto042510,Castellanos083509}%
. This interest is driven essentially by two important properties of these
amplifiers: (1) Their capability to amplify very weak coherent signals; (2)
Their ability to squeeze noise below the equilibrium level by means of
employing a homodyne setup and phase control. These properties are expected to
be highly beneficial to the area of quantum communication
\cite{Houck328,Hofheinz310} and to the generation of quantum squeezed states
\cite{Yurke2519,Movshovich1419,Yurke5054,CastellanosnaturePhys}. Additional
interest in these high-gain parametric amplifiers arises from the large body
of theoretical work predicting photon-generation from vacuum via the dynamical
Casimir effect \cite{BartonS1,Dodonov309}, which can be achieved by employing
an appropriate parametric excitation mechanism \cite{Segev202,Dodonov08064035}%
.%
\begin{figure}
[ptb]
\begin{center}
\includegraphics[
natheight=9.916800in,
natwidth=9.541500in,
height=3.2759in,
width=3.1548in
]%
{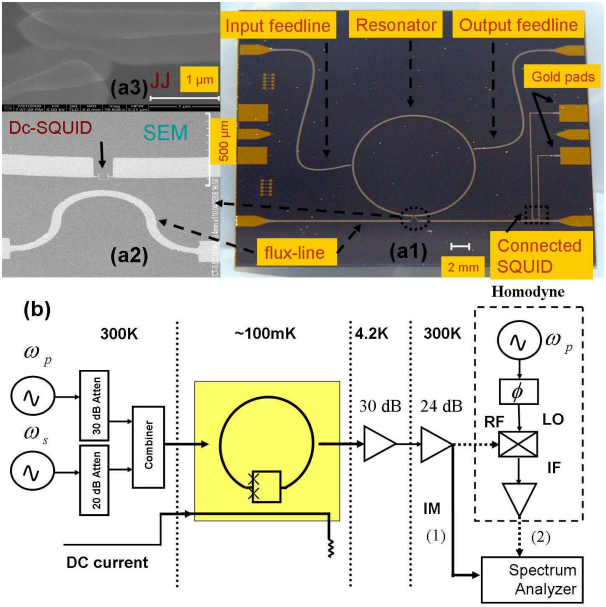}%
\caption{(Color) The device and IM setup. (a1) A photograph of the device
consisting of a circular resonator with a circumference of
$36.4\operatorname{mm}$ and a line width of $150\operatorname{\mu m}$, a
dc-SQUID integrated into the resonator and a flux-line employed for driving
rf-power and flux-biasing the dc-SQUID. (a2) An electron micrograph displaying
the dc-SQUID incorporated in the resonator and the adjacent flux-line. The
area of the dc-SQUID fabricated is $39\operatorname{\mu m}$ $\times$
$39\operatorname{\mu m}$, while the area of each junction is about
$0.25(\operatorname{\mu m})^{2}$. The self-inductance of the dc-SQUID loop is
$L_{s}=1.3\cdot10^{-10}\operatorname{H}$ and its total critical Josephson
current is about $2I_{c}=28\operatorname{\mu A}$. (a3) A micrograph of one of
the dc-SQUID junctions. (b) A basic IM setup. The pump and the signal
designated by their angular frequency $\omega_{p}$ and $\omega_{s}$
respectively are combined by a power combiner and fed to the resonator. The
field at the output is amplified using two amplification stages and measured
using a spectrum analyzer (path 1). A dc current was applied to the flux-line
in order to flux-bias the dc-SQUID. Path (2) at the output corresponds to a
homodyne setup employed in measuring IM.}%
\label{TheDeviceEps}%
\end{center}
\end{figure}

In this present work we study a superconducting stripline microwave resonator
integrated with a dc-SQUID \cite{Sandberg203501,Laloy1034,Nation104516}. The
paper is mainly devoted to a novel amplification mechanism in which the
relatively large nonlinear inductance of the dc-SQUID is exploited to achieve
large gain in an intermodulation (IM) configuration. We provide theoretical
evidence to support our hypothesis that the underlying mechanism responsible
for the large observed gain is metastability of the dc-SQUID. In addition, at
the end of the paper we demonstrate another amplification mechanism, in which
the dependence of the dc-SQUID inductance on external magnetic field is
exploited to achieve parametric gain \cite{Yamamoto042510}.

The device (see Fig. \ref{TheDeviceEps} (a)) is implemented on a high-resistivity $34%
\operatorname{mm}%
\times30%
\operatorname{mm}%
\times0.5%
\operatorname{mm}%
$ Silicon wafer coated with a $100%
\operatorname{nm}%
$ thick layer of SiN. As a preliminary step, thick gold pads ($300%
\operatorname{nm}%
$) are realized at the periphery of the wafer. Subsequently, e-beam
lithography is applied in which both the macroscopic resonator and the
microscopic Josephson gaps are written. Following a developing stage, a two
angle shadow evaporation of Aluminium \cite{Fulton109} -with an intermediate
stage of oxidation- implements the resonator as well as the two Al/AlO$_{x}%
$/Al Josephson junctions comprising the dc-SQUID. The total thickness of the
Aluminium evaporated is $80%
\operatorname{nm}%
$ ($40%
\operatorname{nm}%
$ at each angle). Finally, a lift-off process concludes the fabrication of the
integrated system.

\begin{figure}
[ptb]
\begin{center}
\includegraphics[
height=2.5694in,
width=3.3624in
]%
{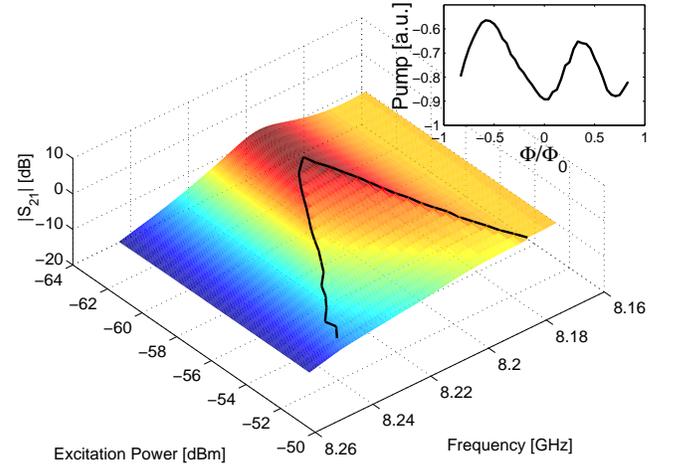}%
\caption{(Color) Transmission response curves of the resonator at its third
resonance ($8.219\operatorname{GHz}$) as a function of input power and
frequency. The resonance curves exhibit nonlinear effects at $P_{p}\geq P_{c}%
$. The loaded quality factor for this resonance in the linear regime is $350$.
The black line maps the jumps appearing at both sides of the resonance curves.
The inset exhibits the dependence of the transmitted pump signal on the
applied magnetic flux threading the dc-SQUID loop during an IM measurement.}%
\label{thirdmoderes}%
\end{center}
\end{figure}

In general, IM generation is a nonlinear phenomenon which is associated often
with the occurrence of nonlinear effects in the resonance curves of a
superconducting resonator
\cite{Dahm2002,Chen4788,Monaco2898,Abdo022508,Tholen253509}. In Fig.
\ref{thirdmoderes} we show a transmission measurement of the resonator
response, obtained using a vector network analyzer at the resonance while
sweeping the frequency upwards. The measured resonance resides at
$f_{0}=8.219$ $%
\operatorname{GHz}%
$ which corresponds to the third resonance mode of the resonator having an
anti-node of the rf-current waveform at the dc-SQUID position. Similar
nonlinear effects in the transmission response have been measured at the first
mode ($\simeq2.766$ $%
\operatorname{GHz}%
$) as well.

As can be seen in the figure, at excitation powers $P_{p}$ below $P_{c}=$
$-60.3$ dBm (at which nonlinear effects emerge) the resonance curve is linear
and Lorentzian. As the input power is increased abrupt jumps appear at both
sides of the resonance (see black line in Fig. \ref{thirdmoderes}) accompanied
by frequency hysteresis loops. In addition, as one continues to increase the
input power the resonance curves become shallower, broader and less
symmetrical. Such power dependency can be attributed to the nonlinearity of
the dc-SQUID inside the resonator which increases considerably as the
amplitude of the driving current becomes comparable to the critical current.%

The basic IM scheme used is schematically depicted in Fig. \ref{TheDeviceEps}
(b). The input field of the resonator is composed of two sinusoidal fields
generated by external microwave synthesizers and superimposed using a power
combiner. The applied signals have unequal amplitudes. One, referred to as the
pump, is an intense sinusoidal field with frequency $f_{p}$, whereas the
other, referred to as the signal, is a small amplitude sinusoidal field with
frequency $f_{p}+\delta$, where $\delta$ represents the frequency offset
between the two signals. Due to the presence of a nonlinear element such as
the dc-SQUID integrated into the resonator, frequency mixing between the pump
and the signal yields an output idler field at frequency $f_{p}-\delta$. Thus
the output field of the resonator, measured by a spectrum analyzer, consists
mainly of three spectral components, the transmitted pump, the transmitted
signal and the generated idler. The IM amplification in the signal, idler and
the noise is obtained, as shown herein, by driving the dc-SQUID to its onset
of instability via tuning the pump power.

\begin{figure}
[ptb]
\begin{center}
\includegraphics[
height=2.5356in,
width=3.4229in
]%
{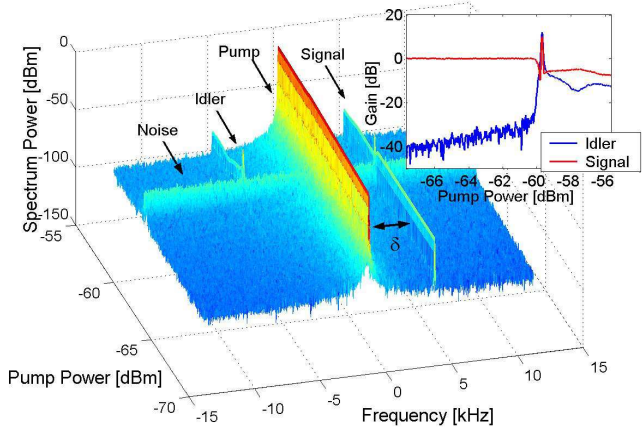}%
\caption{(Color) A spectrum power measured by a spectrum analyzer during IM
operation as a function of increasing pump power while applying external flux
$\Phi\simeq\Phi_{0}/4$. The spectrum taken at a constant frequency span around
$f_{p}$ was shifted to dc for clarity. At the vicinity of $P_{c}$ the system
exhibits a large amplification. A cross-section of the measurement taken along
the pump power axis is shown in the inset. The red line (light grey) and the
blue line (dark grey) exhibit the corresponding gain factors of the
transmitted signal and the idler respectively. }%
\label{ImProductsMesh}%
\end{center}
\end{figure}

In Fig. \ref{ImProductsMesh} we show a typical IM measurement result. In this
measurement the pump was tuned to the resonance frequency $f_{p}=8.219%
\operatorname{GHz}%
$. The signal power $P_{s}$ was set to $-100$ dBm, and its frequency offset
$\delta$ to $5%
\operatorname{kHz}%
$. As the pump power injected into the resonator is increased, the
nonlinearity of the dc-SQUID increases and consequently the frequency mixing
between the pump and the signal increases as well. At about $P_{c}$ the
dc-SQUID reaches a critical point at which the idler, the signal and also the
noise floor level (within the frequency bandwidth) undergo a large
simultaneous amplification. The idler gain measured relative to the injected
signal power at the resonator input is about $13$ dB (see inset of Fig.
\ref{ImProductsMesh}). The same result is obtained as well, as one calculates
the power ratio of the generated signal at the output port of the resonator to
the input signal while accounting for the losses and amplifications of the
elements along each direction.%

In order to show that this IM amplifier is also phase sensitive we applied a
standard homodyne detection scheme as schematically depicted in Fig.
\ref{TheDeviceEps} (see path (2)), in which the phase difference between the
pump and the local oscillator (LO) at the output -having the same frequency-
can be varied. In such a scheme the pump is down-converted to dc, whereas both
signal and idler are down-converted to the same frequency $\delta$. The
largest amplification is obtained when the LO phase ($\phi_{LO}$) is adjusted
such that the signal and idler tones constructively interfere. Shifting
$\phi_{LO}$ by $\pi/2$ from the point of largest amplification results in
destructive interference, which in turn leads to the largest de-amplification.

\begin{figure}
[ptb]
\begin{center}
\includegraphics[
height=2.373in,
width=3.1574in
]%
{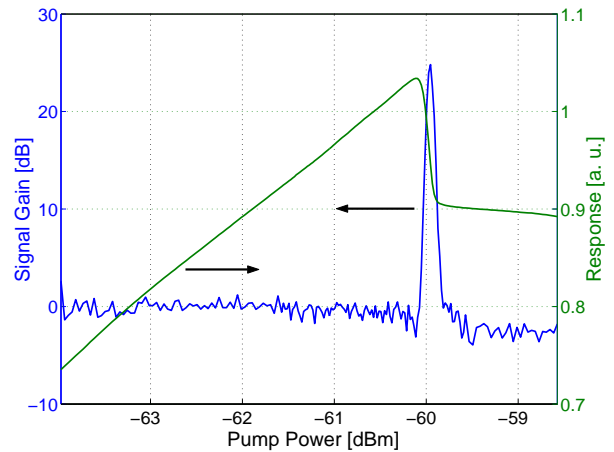}%
\caption{(Color) A signal gain (blue line) measured as a function of
increasing pump power. The measurement was taken using a homodyne scheme. The
signal displays a gain of about $25$ dB at $P_{c}$. The corresponding dc
component of the homodyne detector output (green line) was measured
simultaneously using a voltage meter in parallel to the spectrum analyzer. In
this measurement $P_{s}=$ $-110$ dBm and $\delta=5\operatorname{kHz}$.}%
\label{SignalGainVsPumpHM}%
\end{center}
\end{figure}

\begin{figure}
[ptb]
\begin{center}
\includegraphics[
height=2.597in,
width=3.2292in
]%
{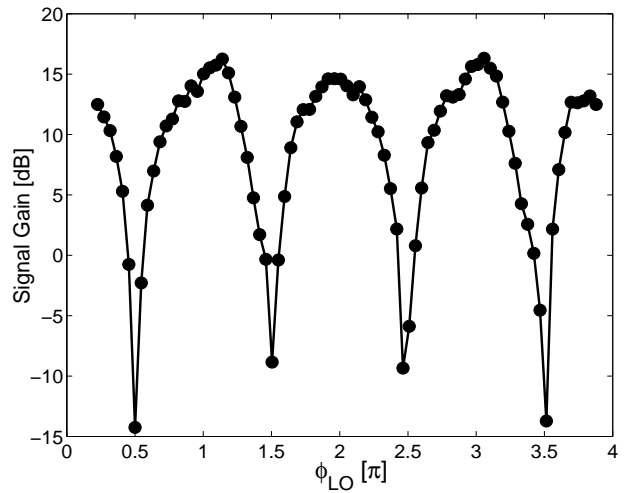}%
\caption{A periodic dependence of the signal gain on the LO phase difference
at the vicinity of $P_{c}$. The signal exhibits a large amplification and
de-amplification at integer and half integer multiples of $\pi$ respectively.
The phase modulation dependency shows up to $30$ dB peak to peak amplitude.}%
\label{signalgainvsphase}%
\end{center}
\end{figure}

The IM measurement results, obtained using the homodyne setup while
flux-biasing the dc-SQUID with about half flux-quantum, are shown in Figs.
\ref{SignalGainVsPumpHM} and \ref{signalgainvsphase}. Figure
\ref{SignalGainVsPumpHM} exhibits the signal gain (blue line) versus
increasing applied pump power. As can be seen from the figure the signal gain
assumes a peak of $25$ dB for $P_{p}\simeq P_{c}$. Moreover, the response of
the resonator at the pump frequency (green line) -measured simultaneously
using a voltage meter connected in parallel to the spectrum analyzer- is drawn
as well for comparison. The sharp drop in the region $P_{p}\sim P_{c}$,
coinciding with the amplification of the signal, indicates that the
transmitted pump is depleted and power is transferred to other frequencies.%

Fig. \ref{signalgainvsphase} exhibits a periodic dependence of the signal gain
on $\phi_{LO}$, having a gain modulation of about $30$ dB peak to peak between
the amplification and de-amplification quadratures. It is worthwhile
mentioning that in this measurement not only the signal shows a periodic
dependence on $\phi_{LO}$ but also the noise floor which exhibits up to $20$
dB modulation (measured at $f_{p}+1.5\delta$).%

\begin{figure}
[ptb]
\begin{center}
\includegraphics[
height=2.6965in,
width=3.2949in
]%
{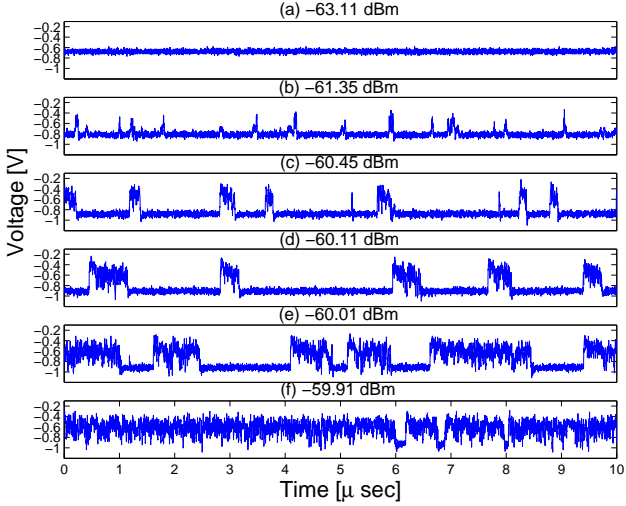}%
\caption{Snapshots of the resonator response measured in the time domain
during an IM operation. (a)-(d) display the resonator response at pump powers
lower than the critical value. (e) At the critical value. (f) Exceeding the
critical value. }%
\label{TimeDomainSnaphShots}%
\end{center}
\end{figure}

Another interesting aspect of this device is revealed by examining its
response in the time domain while applying an IM measurement using the
homodyne setup shown in Fig. \ref{TheDeviceEps} (see path 2). This is achieved by
connecting a fast oscilloscope in parallel to the spectrum analyzer at the
output. A few representative snapshots of the device behavior measured in the
time domain are shown in Fig. \ref{TimeDomainSnaphShots}, where each subplot
corresponds to a different applied pump power. In subplot (a) the pump power
is about $3$ dBm lower than the critical value and it displays a constant
voltage level. As the pump power is increased (subplot (b)) separated spikes
start to emerge. Increasing the power further causes the spikes to start
bunching with one another and as a consequence forming larger groups (subplots
(c) and (d)). This bunching process reaches an optimal point at the pump power
corresponding to the peak in the IM gain (subplot (e)). Above that value, such
as the case in subplot (f), the spikes become saturated and the gain drops.

Such behavior in the time domain reveals also the underlying mechanism
responsible for the amplification. As it is clear from the measurement traces
the rate of spikes strongly depends on $P_{p}$ at the vicinity of $P_{c}$,
thus adding a small signal at $f_{p}+\delta$ (which is effectively equivalent
to applying an amplitude modulation of the pump at frequency $\delta$),
results in modulation of the rate of spikes at the same frequency, and
consequently yields a large response at the signal and the idler tones.%

In an attempt to account for the unique temporal response presented in Fig.
\ref{TimeDomainSnaphShots}, we refer to the two equations of motion governing
the dc-SQUID system given by \cite{Koch380}%

\begin{align}
\beta_{C}\frac{\overset{\cdot\cdot}{\delta}_{1}}{I_{c}}\left(  \frac{\Phi_{0}%
}{2\pi R_{J}}\right)  ^{2}+\overset{\cdot}{\delta}_{1}\left(  \frac{\Phi_{0}%
}{2\pi R_{J}}\right)   &  =-\frac{\partial U}{\partial\delta_{1}%
}+I_{\mathrm{n}1}\nonumber\\
\beta_{C}\frac{\overset{\cdot\cdot}{\delta}_{2}}{I_{c}}\left(  \frac{\Phi_{0}%
}{2\pi R_{J}}\right)  ^{2}+\overset{\cdot}{\delta}_{2}\left(  \frac{\Phi_{0}%
}{2\pi R_{J}}\right)   &  =-\frac{\partial U}{\partial\delta_{2}%
}+I_{\mathrm{n}2}\;, \label{isolatedDCSQUID}%
\end{align}
where $\Phi_{0}$ is the flux quantum, $R_{J}$ is the shunt resistance of the
junctions,$\ I_{c}$ is the critical current of each junction, $\delta_{1},$
$\delta_{2}$ are the gauge invariant phase differences across junctions 1 and
2 respectively, $I_{n1},I_{n2}$ are equilibrium noise currents generated in
the shunt resistors having -in the limit of high temperature- a spectral
density of $S_{I_{\text{\textrm{n}}}}=4k_{B}T/R_{J}$, $\beta_{C}$ is a
dimensionless parameter defined as $\beta_{C}\equiv2\pi I_{c}R_{J}^{2}%
C_{J}/\Phi_{0}$, where $C_{J}$ is the junction capacitance and $U$ is the
potential energy of the system which reads%

\begin{align}
U  &  =-\frac{I}{2}\left(  \delta_{1}+\delta_{2}\right)  +\frac{2I_{c}}%
{\pi\beta_{L}}\left(  \frac{\delta_{1}-\delta_{2}}{2}-\frac{\pi\Phi}{\Phi_{0}%
}\right)  ^{2}\nonumber\\
&  -I_{c}\left(  \cos\delta_{1}+\cos\delta_{2}\right)  \;, \label{Usq}%
\end{align}
where $\Phi$ is the applied magnetic flux, $I$ is the bias current flowing
through the dc-SQUID, and $\beta_{L}$ is a dimensionless parameter defined as
$\beta_{L}\equiv2L_{s}I_{c}/\Phi_{0}$. While the circulating current flowing
in the dc-SQUID loop is given by $I_{\mathrm{circ}}\left(  t\right)
=I_{c}\left(  \delta_{1}-\delta_{2}-2\pi\Phi/\Phi_{0}\right)  /\pi\beta_{L}$.

Furthermore, in order to account for the resonator response as well, we make
two simplifying assumptions: (1) We model our resonator as a series RLC tank
oscillator characterized by an angular frequency $\omega_{0}=1/\sqrt{LC}%
=2\pi\cdot8.219%
\operatorname{GHz}%
$, and a characteristic impedance $Z_{0}=\sqrt{L/C} $, where $L$ and $C$ are
the inductance and the capacitance of the resonant circuit respectively; (2)
We neglect the mutual inductance that may exist between the inductor and the dc-SQUID.

Under these simplifying assumptions one can write the following equation of
motion for the charge on the capacitor denoted by $q\left(  t\right)  $%

\begin{equation}
\frac{Z_{0}\ddot{q}}{\omega_{0}}+R\dot{q}+\omega_{0}Z_{0}q+V_{sq}%
+V_{\mathrm{in}}+V_{\mathrm{n}}=0, \label{RLCeq}%
\end{equation}
where $V_{\mathrm{n}}$ is a voltage noise term having -in the limit of high
temperature- a spectral density of $S_{V_{\mathrm{n}}}=4Rk_{B}T$,
$V_{\mathrm{in}}\left(  t\right)  =\operatorname{Re}\left(  V_{0}%
e^{-i\omega_{p}t}\right)  $ is an externally applied sinusoidal voltage
oscillating at the pump angular frequency $\omega_{p}$ and having a constant
voltage amplitude $V_{0}$. Whereas, $V_{sq}=\Phi_{0}\left(  \dot{\delta}%
_{1}+\dot{\delta}_{2}\right)  /4\pi$ designates the voltage across the dc
SQUID. Using these notations the bias current in Eq. (\ref{Usq}) reads
$I=\dot{q}$.

Finally in order to relate the observed output signal in Fig.
\ref{TimeDomainSnaphShots} at the output of the homodyne scheme to the fast
oscillating solution $q$, we first express the charge on the capacitor as
$q\left(  t\right)  =\sqrt{2\hbar/Z_{0}}\operatorname{Re}\left[  A\left(
t\right)  e^{-i\omega_{p}t}\right]  $, where $A\left(  t\right)  $ is a slow
envelope function, second, we employ the following input-output relation
\cite{Yurke5054} $V_{\mathrm{out}}\left(  t\right)  =V_{0}-i\sqrt{32Z_{0}%
\hbar\gamma_{1}^{2}}A\left(  t\right)  $, where $\gamma_{1}$ is the coupling
constant between the resonator and the feedline, in order to obtain the field
at the output port. third, we account for the phase shift by evaluating the
following expression $2V_{LO}\operatorname{Re}\left[  V_{\mathrm{out}}\left(
t\right)  e^{i\phi_{LO}}\right]  $, where $V_{LO}$ corresponds to the
amplitude of the LO.

\begin{figure}
[ptb]
\begin{center}
\includegraphics[
height=2.7605in,
width=3.3615in
]%
{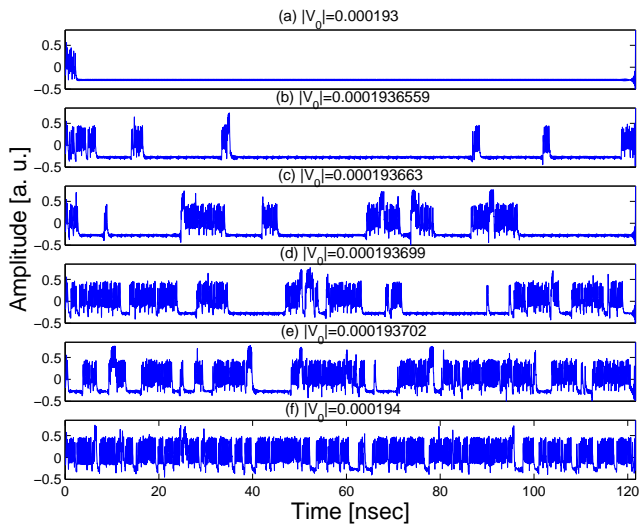}%
\caption{A simulation result. The output field exhibits spikes in the time
domain similar to the spikes shown in Fig. \ref{TimeDomainSnaphShots}. The
spikes occur whenever the system locus jumps from one well to another due to
the presence of stochastic noise while driving the system near a critical
point. In this simulation run the system was simulated over a period of $1000$
time cycles of pump oscillations (one of the constraints set on the simulation
was maintaining a reasonable computation time). The parameters that were
employed in the simulation (same as experiment) are $\omega_{0}=\omega
_{p}=2\pi\cdot8.219\operatorname{GHz}$, $\beta_{L}=3.99$, $R_{J}%
=9.4\operatorname{\Omega }$, $\Phi=\Phi_{0}/2$ and $\phi_{LO}=\pi$. The rest
of the parameters were set in order to reproduce the measurement result:
$\gamma_{1}=10^{7}\operatorname{Hz}$, $R=3.3$ $\operatorname{\Omega }$,
$\beta_{C}=5.86$ and $Z_{0}=10\operatorname{\Omega }$ (corresponding
experimental values are: $\gamma_{1}\simeq2.4\cdot10^{7}\operatorname{Hz}$,
$Z_{0}\simeq50\operatorname{\Omega }$, $C_{J}$ was not measured directly, a
capacitance on the order of $0.7\operatorname{pF}$ was assumed).}%
\label{SimulationResult}%
\end{center}
\end{figure}

By integrating these stochastic coupled equations of motion numerically, while
employing device parameters which are relevant for our case, one finds that
the observed temporal behavior of the system can be qualitatively explained in
terms of noise-induced jumps between different potential wells forming the
potential landscape of the dc-SQUID which is given by Eq. (\ref{Usq}). As a
consequence of applying the voltage $V_{\mathrm{in}}$ to the integrated
system, the potential landscape of the dc-SQUID oscillates at the pump
frequency, and its oscillation amplitude grows with the amplitude of the
incoming voltage. However, inter-well transitions of the system become most
dominant as the oscillation voltage reaches a critical value at which the
potential landscape of dc-SQUID becomes tilted enough -due to the current
flowing in the system- in order to allow frequent noise-assisted escape events
from one well to another or across several wells, which in turn cause a
voltage drop to develop across the dc-SQUID and induce jumps in the
circulating current. Such hopping of the system state is manifested as well in
the homodyne output field response shown in Fig. \ref{SimulationResult}, which
in a qualitative manner mimics successfully the main temporal features shown
in Fig. \ref{TimeDomainSnaphShots}.

It is also clear from this inter-well transitions model that the amplification
in this case is different from the so-called Josephson bifurcation amplifier
\cite{Siddiqi207002}, as in the latter case the system is confined to only one
well and the bifurcation occurs between two oscillation states having
different amplitudes.%

\begin{figure}
[ptb]
\begin{center}
\includegraphics[
natheight=12.562300in,
natwidth=9.906400in,
height=3.7178in,
width=2.9378in
]%
{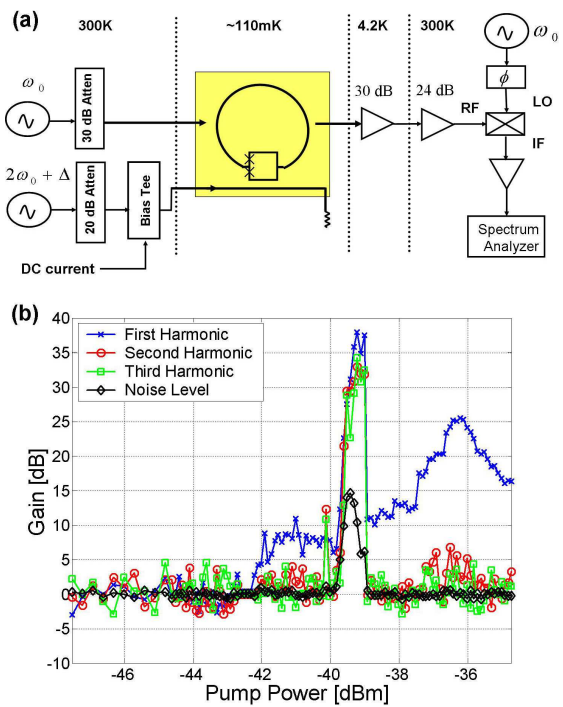}%
\caption{(Color) (a) A homodyne setup employed in measuring parametric
excitation. (b) Large gain measured in a parametric excitation experiment
corresponding to the generated three harmonics (at $\delta=5\operatorname{kHz}%
$, $2\delta$, $3\delta$ respectively) and the noise spectral density at
$1.5\delta$. The gain was measured as a function of increasing pump power
applied at $2f_{0}+\delta$. The signal power applied to the resonator at
$f_{0}$ was $P_{s}=-80$ dBm. In this measurement a maximum gain of $38$ dB is
achieved at the first harmonic. }%
\label{PEHmSetup}%
\end{center}
\end{figure}

Moreover, just as in the recent experiment by Yamamoto \textit{et al.}
\cite{Yamamoto042510} we have additionally employed our device as a parametric
amplifier. To this end, we have used the parametric excitation scheme
exhibited in Fig. \ref{PEHmSetup} (a) in which the pump and signal tones are
applied to different ports. The main rf signal (pump) is applied to the
flux-line at the vicinity of twice the resonance frequency of the resonator
$2f_{0}+\delta$ ($\delta\equiv\Delta/2\pi=5%
\operatorname{kHz}%
$) and $2f_{0}=16.438%
\operatorname{GHz}%
$ does not coincide with any resonance of the device. Whereas, the signal,
being several orders of magnitude lower than the pump, is fed to the resonator
port at $f_{0}$ and its main purpose is to probe the system response.%

In Fig. \ref{PEHmSetup} (b) we exhibit a parametric excitation measurement
result obtained using this device at $\Phi\simeq0.6\Phi_{0}$. In this result
there is evidence of one of the characteristic fingerprints of a parametric
amplifier: the existence of an excitation threshold above which there is a
noise rise and an abrupt amplification of the harmonic at $f_{0}+\delta$
(which results from a nonlinear frequency mixing at the dc-SQUID). As can be
seen in the figure at about $-40$ dBm the first harmonic generated at $\delta$
and the two higher-order harmonics at $2\delta$, $3\delta$ get amplified
considerably up to a maximum gain of $38$ dB measured at the first harmonic.
Also in a separate measurement result (not shown here) the first harmonic has
been found to display $\simeq20$ dB peak to peak modulation as a function of
external magnetic field.

In conclusion, in this work we have designed and fabricated a superconducting
stripline resonator containing a dc-SQUID. We have shown that this integrated
system can serve as a phase sensitive amplifier. We have studied the device
using IM measurement and parametric excitation. In both schemes the device
exhibited distinct threshold behavior, strong noise rise and large
amplification of coherent side-band signals generated due to the nonlinearity
of the dc-SQUID. In addition, we have investigated the system response in the
time domain during IM measurements. We have found that in the vicinity of the
critical input power the system becomes metastable and consequently exhibits
noise-activated spikes in the transmitted power. We have shown that this kind
of behavior can be explained in terms of noise-assisted hopping of the system
state between different potential wells. We have also demonstrated that the
main features observed in the time domain can be qualitatively captured by
solving the equations of motion for the dc-SQUID in the presence of rf-current
bias and stochastic noise. Such a device may be exploited under suitable
conditions in a variety of intriguing applications ranging from generating
quantum squeezed states to parametric excitation of zero-point fluctuations of
the vacuum.

\begin{center}
\textbf{ACKNOWLEDGEMENTS}
\end{center}

This work was supported by the Israel Science Foundation, the Deborah
Foundation, the Poznanski Foundation, Russel Berrie Nanotechnology Institute,
US-Israel Binational Science Foundation and MAFAT. BA was supported by the
Ministry of Science, Culture and Sports.

\bibliographystyle{plain}
\bibliography{attachfile}

\end{document}